# Observation of Lightning Ball (Ball Lightning): A new phenomenological description of the phenomenon

Domokos **Tar**

M.dg. in physics of: Swiss Federal Institute of Technology, ETH-Zürich, Switzerland
CH-8712 Stäfa/Switzerland, Eichtlenstr.16,

**Abstract**

The author (physicist) has observed the very strange, beautiful and frightening Lightning Ball (LB). He has never forgotten this phenomenon. During his working life he could not devote himself to the problem of LB-formation. Only two years ago as he has been reading different unbelievable models of LB-formation, he decided to work on this problem.
By studying the literature and the crucial points of his observation the author succeeded in creating a completely new model of Lightning Ball (LB) and Ball Lightning (BL)-formation based on the symmetry breaking of the hydrodynamic vortex ring. This agrees fully with the observation and overcomes the shortcomings of current models for LB formation. This model provides answers to the questions: Why are LBs so rarely observed, why do BLs in rare cases have such a high energy and how can we generate LB in the laboratory? Moreover, the author *differentiates between LB and BL*, the latter having a high energy and occurring in 5 % of the observations.

## 1. Observation, an eyewitness report.

In the year 1954 I was a physics student a the University of Budapest in my 4-th semester. On a warm sommer day at about 10 am I was walking on Margaret Island which is situated between 2 branches of the Danube (Fig.1). It is a nature park with a large meadow. The temperature was about 25-27 °C and it was very muggy.The sky was already covered with dark clouds. A thunderstorm was approaching, distant thunder was heard. The wind and rain began. The front approached very quickly. There was absolutely no shelter nearby. The humidity was about 100% because of the rain. Suddenly direct before me a terrible flash of lightning struck the ground in full view. Such a loudness I have never heard in my life. The lightning channel was about 50 m away from me, directly in my line of vision, so I did not need to turn my head to be able to see the lightning. Hence I could observe in detail the whole phenomenon. The channel was very brilliant and had a diameter of about 25-35 cm. It was a direct verticle straight line to the ground. The lightning flash illuminated a bush (with trees in the centre) which was on the right at a distance of about 2.5 m from the stroke grounding point (Fig.3). This 2 m high bush was the only one on the large grassy surface. Immediately a very strong wind began to blow. I saw the bushes bend to the right under the wind. After the lightning, it got relatively dark, because of the dark clouds. In the background there was a dark building. The bushes were still moving in the wind and on the left wet leaves and grasses were whirling through the air (Figure 2). After about 2 sec of darkness, suddenly a very beautiful sphere with a diameter of about 30-40-cm appeared about 1.2 m above the ground, at the same distance (about 2.5 m) from the lightning impact point as the bush, but in the opposite direction, to the left (Fig.2 and 3). The Lightning Ball (LB) was very brilliant, like a little sun. It was spinning counter clockwise. The axis of the rotation was horizontal to the ground and perpendicular to a straight line drawn between the bush (trees), impact point and the ball. The LB had one or two plumes or tails (Fig.5). The plume was not as brilliant as the sphere, and reddish. Very strange was that the plume was not on a perpendicular plane to the ground, but to the north of this plane. In other words, the plume had a component in the YZ-plane. After a very short time about 0.3 sec the plume merged in the ball. The LB was moving at slow, constant speed to the left on straight line mentioned. It had absolutely sharp contours. Its brightness was constant across the entire surface (Fig.4). At that moment, I no longer saw any rotation. My first thought was: *"What strange phenomenon exist in Nature!"* It moved horizontally at constant height and constant speed. It moved along the straight line that could be drawn from the center of the bush the impact point and beyond. I could not observe any change in its diameter. After about 3 sec, the LB disappeared very abruptly, like a soap bubble (Fig.4). I did not hear any noise when it disappeared, which may have been because of the wind and rain [1, 2].

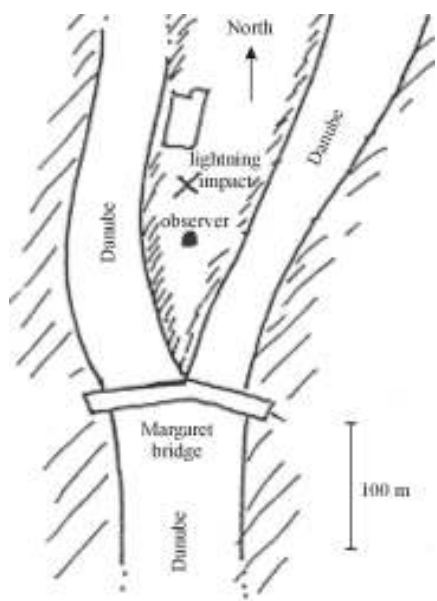

**Schematic of the site where the phenomenon occured**

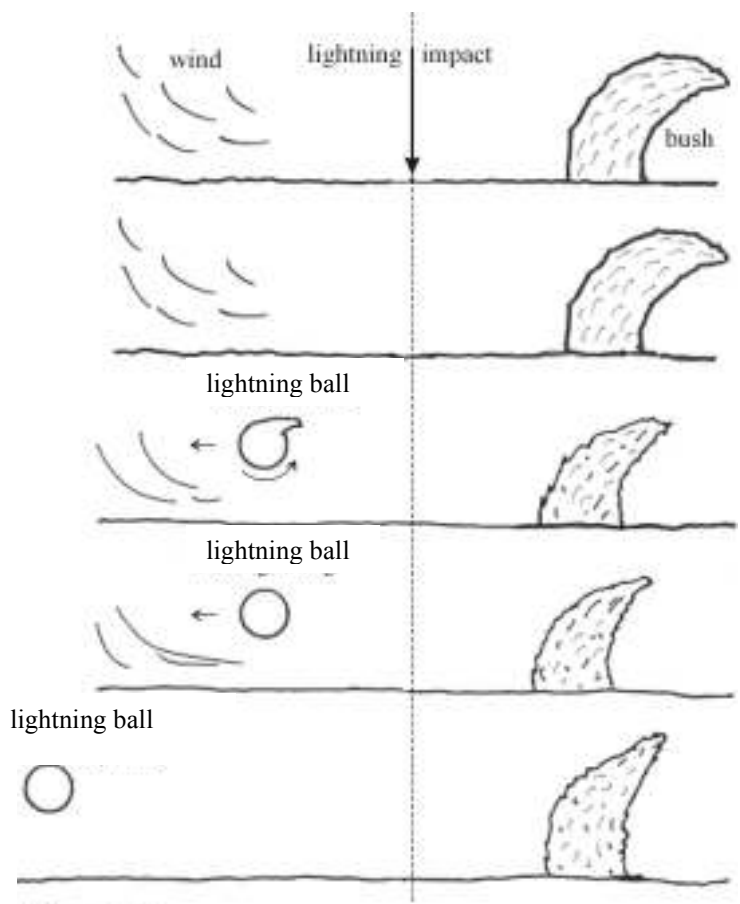

Fig.2: **Sequence of events during observation of the lightning ball**

## 2. The crucial points of the observation

1) The lightning channel is perpendicular to a large plane. An abrupt change in the velocity of the hot channel, that means a stop, is ideal for the creation of the vortex ring.
2) The centre of the bush, the channel impact point and the points of the ball's appearance and disappearance are all on the same straight line (Fig.3).
3) The rotation axis of the remaining cylinder-part of the vortex ring is horizontal to the ground and perpendicular to the straight line in Figure 3, which is represented by the X- axis in Figure 5.
4) The rotation axis of the LB is the same as the rotation axis of the remaining cylinder-part of the vortex ring and is in the Y-axis in the Figure 5.
5) The spinning direction of the ball is counter clockwise (seen from the observer and the lightning impact position). Therefore also the spin of the vortex ring at its origin (near the channel in Figure 6) is the same. That corresponds well with the general physical fact of hot air flowing upwards near the lightning axis while cold air flows along the ground from outside to the channel (Fig. 2 and 6).
6) A very important conclusion based on the revolving direction is that the vortex ring is not part of the lightning channel; otherwise it would turn in the opposite direction as a hydrodynamic continuous laminar flow [14]. It does not mix with the channel either, but exists on his own. Moreover, it does not receive energy from the channel, therefore it is rather cold.
7) The translations velocity of the ball is the same as the spreading- out velocity of the vortex ring. (The balls speed=3.5m/3s=1.17 m/s and this is equal to the speed of expansion of the invisible vortex ring 2.5 m/2s=1.25 m/s. Both velocities are nearly the same, compare Fig.3 with Fig.4.

2(10

8) The plume of the ball shows up not only in the XZ-plane, but also in the XY-plane too (Figure 5). This is an indication that the plume is nothing other than a small remaining part of the big contracted vortex ring. The vortex ring shrank into the ball and its plume. This we call Lightning Ball (LB).

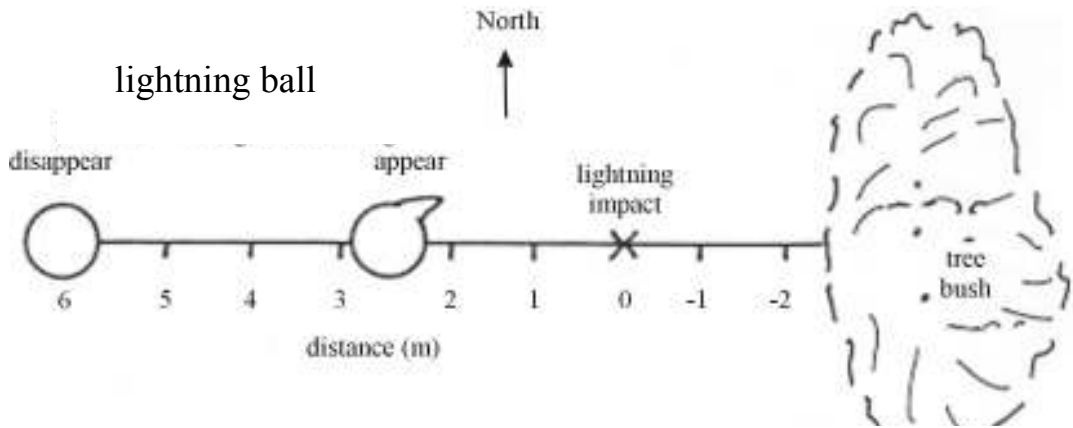

Figure 3

**The path of the lightning ball (seen from above)**

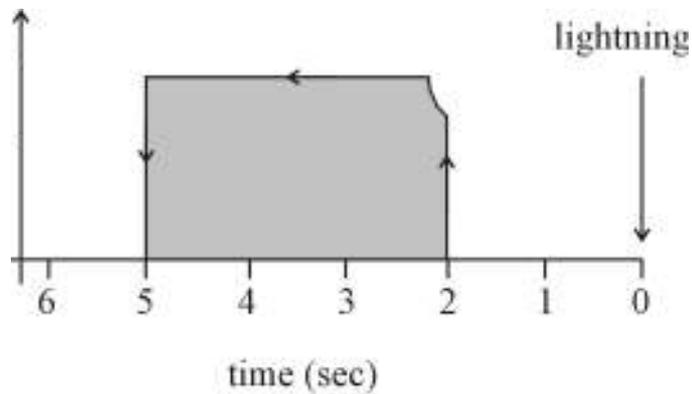

Figure 4

**The brightness of the lightning ball as a function of time**

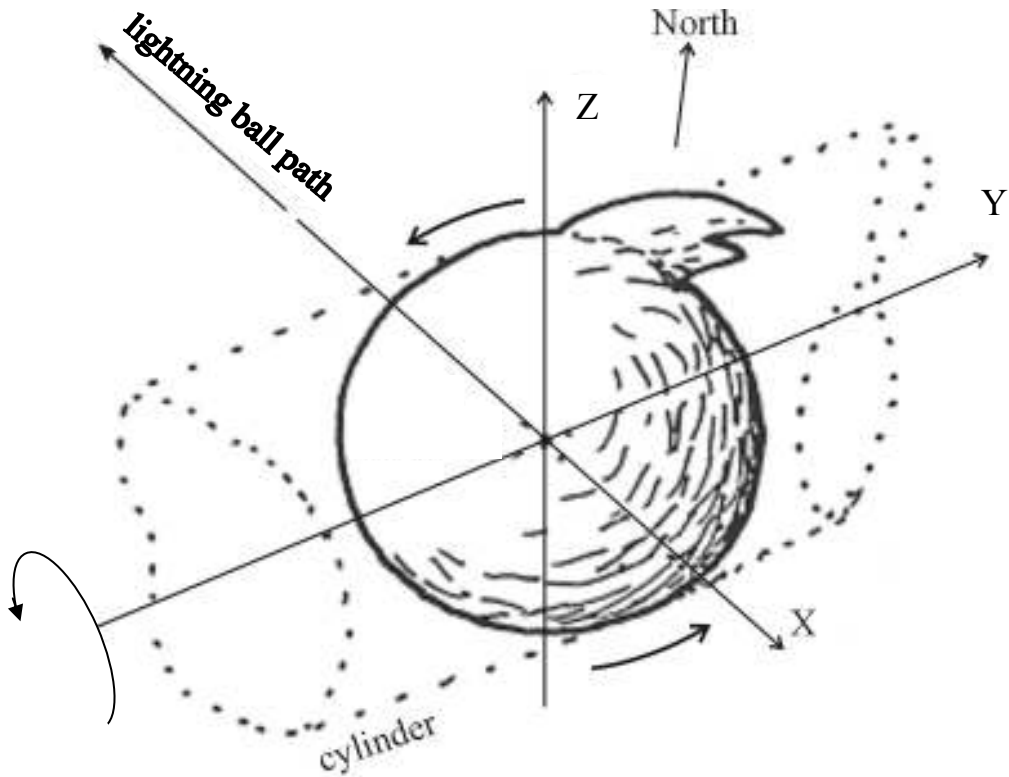

**Appearance of the lightning ball with its plume**

### 3. What we know about the lightning channel and the thunder

In references [3-10], a lot of material can be found on BL and the lightning channel. From spectroscopic measurements we know that the temperature of the lightning channel is 20-30 k °C [12]. The clouds have a main electric potential of 20-50 MV to the ground. The channel current measures between 10 and 100 kA. The first discharge strikes in 10-20 μsec, but it happens in several steps. In general, there are further return strikes in a short time. The magnetic field around the lightning channel vanishes in about 40 msec. We know that lightning also produces high-frequency (HF) electromagnetic fields, but they disappear in a very short time of 50 μsec. The lightning energy can reach 10 MJoule [10]. Of the initial discharge energy practically the entire amount goes into shockwave and sound energy (thunder), that is about 99.99 %. The first shockwave travels much faster than sound in air. The high speed shockwave travels only in the neighbourhood of the lightning channel, than thunder follows at normal speed. First of all, there is an explosion above and implosion near the ground. This determines the rotation direction of the vortex ring around its ring axis. At a laminar flow this spin is just in the opposite direction.

# Theory

### 4. Description of the formation of the vortex ring

The vortex ring is a low pressure ring caused by the lightning channel, thunder and by the whirlwinds (Fig.6.). Its energy is about $10^5$ times lower than the lightning's discharge energy. It resembles a tyre that is pumped up. It turns on its own ring axis. It is not identical with the HILL's vortex ring [12-14]. Different forms of the hydrodynamic vortex rings are described in the work of Kopiev [15]. If there is a large horizontal plane and there is nothing in the neighbourhood of the channel, the ring grow and broaden. This is the ideal case of expansion. A

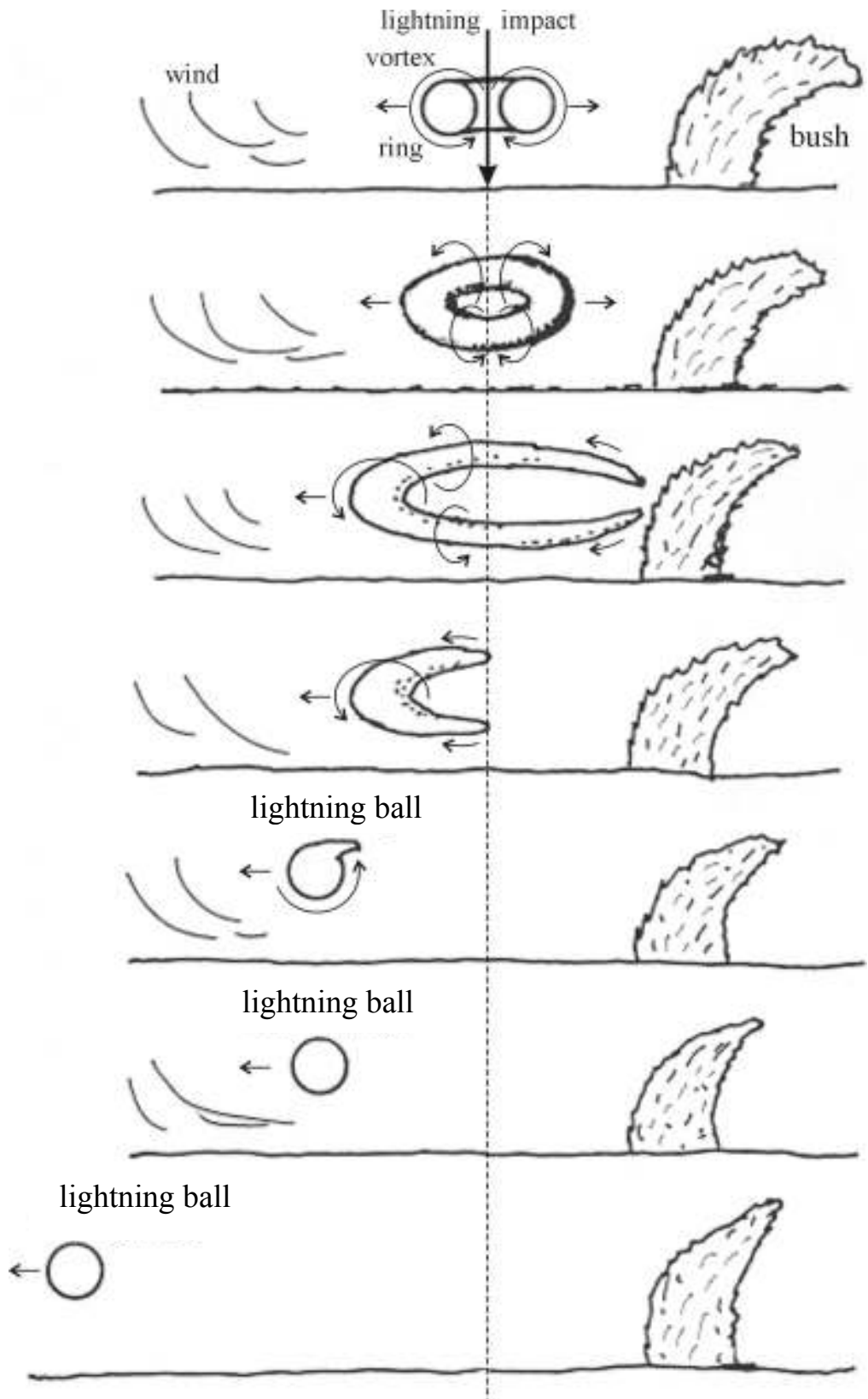

## Figure 6

**Theory of the formation of the lightning ball from a hydrodynamic vortex ring**

treatment of urolithiasis. A high voltage, high current discharge pulse from a generator (a little thunder) is used to destroy kidney stones. Here too a vortex ring forms, but of small dimensions. Photographs of these can be seen in references [16, 17].

- The vortex ring could not be observed because it was cold and the molecules had not yet been excited and therefore the ring did not radiate visibly. This happened later by greater electrification of the enclosed gases and consequently the sphere with its plume lit up and became observable to the human eye.
- The probable raison why the vortex ring could expand further was the bush and so the ring broke upon reaching it (Fig.6), as also the time and distance elements indicate this (Figures 3 and 4). After breaking the ring shrank rapidly into a sickle form (still invisible) and subsequently into a more stable and visible **Lightning Ball (LB).**

## 5. Internal energy through triboelectrification and the rise of a central force

Lenard proved [18] that at waterfalls a huge quantity of charged aerosol particles is created [19, 20]. The more violent the convection and turbulence, the stronger the electrification. This phenomenon is called triboelectricity. The negative charged particles created represent a high quantity of very electrons with high mobility. The heavy ions with low mobility carry the positive charges. They automatically stay in the centre of the rotating cylinder and later in the LB sphere. The negative electrons accumulate on the surface of the ball. Exterior charges can influence the ball's surface, creating a double layer [19, 21]. After the creation of the ball, a defined stable revolution exists only for a certain time. After that the LB floats with no defined rotation axis. This we call the Lightning Ball (LB). In the author's observation the ball appeared only 2 sec after the flash (Fig.4). This is proof that the vortex ring at its creation was rather cold, because about 2 sec were needed to deliver enough energy for the necessary excitation of water drops and gas molecules to metastable molecular electronic states by triboelectrification and electroluminescence so that a lightning ball event could occur. The triboelectricity of the atmosphere is the result of the continuous movements of the water clouds. It is also the source of LB energy. The main part of lightning energy is transferred into the shock and sound energy (99.99%). Only a very small part goes in the LB [10]. The ball's brightness may have been similar to an opaque electric white light bulb of about 200-800 W intensity. On the assumption that the ball is a spherical electrical capacitor loaded on the breakdown voltage, the energy can be calculated by $W = C \cdot U^2 / 2$, where $C = 4 \cdot \pi \cdot \varepsilon_o \cdot R$ is the capacity and R the ball radius = 0.15 m; $U = 1.2 \cdot 10^6$ V/m is the breakdown voltage in air and $\varepsilon_0$ is the influence constant of $9 \cdot 10^{-12}$ As/Vm. For the energy we obtain only 12 Joule, which is very small. The very white and bright appearance may stem from a high water-vapour content or the highly excited electroluminescence of $O_2$, $N_2$, $CO_2$, $H_2O$ molecules. The density of the ball must have been equal to the density of air. The entire ball was probably neutral. But this does not mean that the ball would not be dangerous for people (a person can be killed by 30 mA current). The lifetime of the ball can vary widely, from seconds to minutes, depending on the energy stored. Different external circumstances can shorten its lifetime. For example, injuries can break the surface layer, which leads to an explosion. There is a difference in the surface tension between a soap-bubble and LB. The LB can change its shape, thus can pass through an opening. According to eyewitnesses, LBs pass mainly through open windows and doors with the flow of air. In one case, LB has even penetrated into aircraft. It is not difficult to understand this phenomenon. It is practically impossible to shield a space completely. That is to say those HF-fields are like gases and can easily pass through small holes of some millimetres, for example, in seals around doors or windows. LBs have an internal HF field, because of the corona discharges and glow discharges similar to St.Elmo's fire. These are also HF phenomena, and help LBs to penetrate into aircraft. However this is the exception.

## 6. The stable path of the Lightning Ball

The vortex ring expands radially, concentrically with a constant speed (Fig.6). Around it there are high speeds whirlwinds but the ring itself is very stable. The LB sphere is a prolongation of the ring's path. The sphere inherits the stability of the vortex ring for the first time of its life and it is guided by the Earth's surface on a horizontal stable path [22].

## 7. Why do BL so rarely emerge from a lightning strike?

Berger a meteorologist and lightning scientist saw and photographed many thousand of lightning flashes on the top of Mont Salvatore (CH) during his 30 years of experience, but he never saw a LB [23, 10]. The probable reasons for this are the following:

1) Lightning discharges naturally search out pointed objects that lie higher than their surroundings, because of the corona discharge (field strength). This is the case almost everywhere, but especially so in the mountains. In this way either the vortex ring cannot develop or can easily spread, broadening out in a great space-angle, without developing a LB (no abrupt change).
2) For a LB to occur the lightning flash should possibly happen vertically and abruptly on a large horizontal surface. This also does not apply in the mountains.
3) Apart from a horizontal plane, the immediate surroundings should be asymmetrical, that means some obstacles that has low electrical conductivity should exist on one side, so that the broken vortex ring can be directed in only one direction.
4) The air humidity, temperature and pressure should be sufficient. In the mountains this is less the case.

The author thinks that the photographs of LB reproduced in the book by Stenhoff [10] on pages 149, 150, 151, and 154 and those in the book by Barry [5] on pages 104, 108, 109 and 111 are reliable.

## 8. The high energy ball-lightning models

The author read about 400 eyewitness reports on LB described in [3-11]. LBs in most cases are harmless (see the calculation above). Only in some cases reported did they cause great damage, with energies up to 10 MJoule [10]. This we call **Ball Lightning** [10]. To solve this contradiction with the phenomenon of LB, different energy models were proposed, but always with the same model for both low and high-energy phenomena. Let us look at these models:

### 8.1 Energy from high-frequency standing waves?

Kapitza proposed that the high energies could be explained by the resonant absorption of HF-standing waves [24]. Measurements showed that in realty during and after lightning strikes, HF fields indeed existed, but they were too weak to explain the large damage [10].

### 8.2 Hydrodynamic vortex ring model due to the channel's energy

According to the model of Nickel the ground end of the lightning channel rolls up and separates itself from the channel, creating its own vortex ring [12]. If this model were correct, we should have immediately observed the LB. In this way the channel's material descending from above carries the high energy into the ball. In this ball there are two completely separated "clusters" that turn in opposite directions. But this is nothing other than a HILL vortex ring. This is a mathematical point solution of the Euler equations under the following assumptions: (a) the gases are incompressible (which is not true) and (b) the gases are frictionless (neither is true). Furthermore the HILL's vortex ring does not have spherical symmetry: it has only two rotations axis, one of which is a circle and therefore it has not enough degree of freedom. The ball in contrast three independent rotations-symmetry axes. Nickel proposed that HILL's vortex is not a solution, but should be regarded as a mathematical approach to the problem. This however does not exist in Nature.

### 8.3 Magnetohydrodynamic vortex ring model due to the channel's energy

These models [25, 26] are quite similar to the model of Nickel [12, 13] in that the entire vortex ring is inside of the ball, but the energy is delivered by magneto hydrodynamic forces in the plasma. This model has the following shortcomings:

- According to spectroscopic measurements, the channel temperature does not exceed 20-30 k°K, which is too low for a magneto hydrodynamic generator [8]. Therefore there is no fusion, no high temperature plasma, and it is very improbable that this would happen later in the BL because of the magnetic pinch effect [25]. In any case no radioactive traces have been found [9, 10, 27].

- The channel's magnetic field disappears in a very short time [28], within milliseconds, because, firstly, the return strike, which always happens in the opposite direction, neutralizes the magnetic field created before. If the magnetic vortex ring model were correct, we should practically always observe BL at the impact and not later.

- Moreover, no explanation of BL should have to invoke a violation of the laws of conservation of mass, momentum and energy.

*Remark of the author:*
No one of the above-mentioned authors of these BL theories has ever seen BL.

## 9. The author's hydrodynamic model for the formation of low energy Lightning Ball

The author's LB model is completely different from the above mentioned BL (vortex ring) models. Because:

1) The author's LB does not contain any vortex ring, in contrast to the other models [12, 25, 26]. The ball develops only from a hydrodynamical vortex ring.
2) The author's LB develops outside of the lightning channel and therefore does not contain material and energy from the channel. The direction of ball's rotation proves this. The balls of the other models are formed out of channel's material.
3) The observed LB turns in the opposite direction to that of an impact of laminar flow to the ground plane [14]. The reason for this is that heated air normally rises close to the channel axis and cold air on the ground streams to and up the channel. An other reason is the expansion above and implosion at the bottom (vacuum bubble cavitation) of the vortex ring.

*Further* developments in the formation of LB: The cold hydrodynamic evacuated vortex ring can break into a sickle form and after that shrinks into a revolving cylinder (Figure 5). This happens almost instantly, thus creating additional excited molecules. Central forces can build up, which will dominate over cylindrical forces and finally the LB appears as an illuminated sphere at low energy. The plume of the ball reveals the recent history, the last part of the rotating cylinder (sickle). This corresponds entirely with observation. In fact, there is a symmetry breaking from a lower to a higher symmetry from the vortex ring to the ball. The ring is very unstable because it has only two symmetry axes (one which is a ring). In contrast, the ball has three independent and perpendicular stable symmetry axes. At last an illuminated sphere in the visible the **LB** appears. This corresponds fully with the observation.

## 10. Author's model of delivering high external energy by the Ball Lightning
### Lightning Ball and Ball Lightning: two different phenomena

Stepanov provided good statistics on high damage until 10 M Joule, attributed to BL all of which happened mainly outside of buildings, where DC electrostatic fields could act [29, 30]. Inside buildings, LBs are well shielded and thus cannot attract energy from outside. According to these statistics, many of these cases of great extensive damage can be explained by *ordinary lightning*. This can easily happen because the eyewitnesses concentrated on the very mysterious LB and they did not pay attention to the normal lightning. That means the reliability of the observers could not have been very high. In any case, the rarely observed BL cases with high energy can not be explained only by their internal energy supply. (The electrostatic energy of a comparable spherical capacitor is only 12 Joule.). In the author's view, this mysterious external energy supply mechanism without visible lightning can be described by a triggering mechanism of DC charges with the help of the preliminary breakdown pulses (PBP) [11]. Even in fair weather conditions there is an atmospheric leakage current of 2 pico A/m2, which stems from the global electrical circuit. It is also known that after thunderstorms the DC fields are disturbed and not only locally: these disturbances can stretch out over long distances. So if we have a LB in seemingly fair weather conditions the whole atmospheric environment for some 50 km must be taken into account. Many investigators [11] discovered, that there is an invisible group of about 10 electromagnetic impulses of 30 microsec duration, called PBR (preliminary breakdown impulses) at a distance of 20 to30 km, appearing about 1 msec before each flash of lightning. The PBPs can trigger the so called "streamers". The streamers are invisible air–ion channels with much higher conductivity than the normal air (forerunner). A lightning ball or other source can send out such PBPs and thus trigger the streamers, which then deliver high energy in the form of DC charges to the LB itself. This is a rather *invisible* charge displacement with an explosion in the LB itself. The entire triggering process happens very quickly, in about 50 msec, and thus the explosion happens immediately after charge displacement to the LB. Therefore the entire energy was attributed to the LB itself, but this phenomenon is in reality a *ball lightning* (BL). Its energy can amount to 10 MJoule, depending on the actually accumulated charges of this part of cloud. The probability of such a triggering with high energy is much lower then the probability of the formation of LB itself with low energy. A photograph of

*Proceedings of the 9th International Symposium on Ball lightning (ISBL-06), 16-19 August 2006, Eindhoven, The Netherlands, Eds. G.C. Dijkhuis, D.K.Callebaut and M.Lu ,pp.222-232.*lightning ball with two flashes coming to the ball (at the right and left side), is shown in the ref. [31], pp.335-338, taken in 1928. This photograph proves well the author's model of BL- formation, that means how can a LB get energy from the invisible or visible streamers, forerunner flashes. A phenomenon similar to a LB was photographed by a young man in Uzwil (CH) in the stormy night of July 2004, a strange circular lightning object [32]. The object showed a lot of concentric circles at variable colours. Specialists had examined the film and concluded, it could have been a lightning ball. In the author's opinion: this was a turning cylinder of air masses (line vortex) created by a shock-wave following a lightning flash. The axis of the cylinder can be determined on the photograph. No vortex ring could have developed because of the wide space angle of expansion. No ideal conditions were present for this. The rotation rings were excited gas molecules like $H_2O$, $O_2$, $CO_2$ and $N_2$ generated by the triboelectricity and emitting in the visible at different wavelengths.

## 11. Recommendations for observing or creating Lightning Ball

To be able to observe a lightning ball, one should first create and observe the life of the vortex ring. The following circumstances should be realized:

1) A large abrupt change of a lightning channel: a strike perpendicular to a large horizontal area. Below the ground, a large metallic grid should be placed.
2) The vortex ring should be broken up and directed in one direction only, which can be achieved by an asymmetrical placing of low-conductivity obstacles at one side.
3) There is a time of about 2 sec for observing the invisible vortex ring. The ring can be observed by using the Schlieren optical method, infrared video cameras and by Laser anemometry. Thus the observation of the vortex and later the LB should be possible.
4) These experiments should be made in the open space.

## 12. Closing remark:   "Nature very unwillingly shows us her beautiful Lightning Ball, only if she cannot find another way out !"

**Figures:**
Figure 1 -  Schematic of the site where the phenomenon occurred
Figure 2 -  Sequence of events during observation of the lightning ball
Figure 3 -  The path of lightning ball (seen from above)
Figure 4 -  The brightness of the lightning ball as a function of time
Figure 5 -  Appearance of the lightning  ball with its plume
Figure 6 -  Theory of the formation of the lightning ball from a hydrodynamic vortex ring